\documentclass[article,twocolumn,
amssymb,aps,prd,groupedaddress,superscriptaddress,nofootinbib]{revtex4-2}
\usepackage{amsmath, amssymb}
\usepackage{hyperref}
\usepackage{graphicx}
\usepackage{bm}
\usepackage{mathrsfs}
\usepackage{enumerate}
\usepackage{tensor}
\usepackage{color}
\usepackage{soul}

\begin{document}

\title{Dark Energy from Time Crystals}

\author{Laura Mersini-Houghton}
\email{mersini@physics.unc.edu}
\affiliation{Department of Physics and Astronomy, UNC-Chapel Hill, NC, USA 27599}

\date{\today}

\begin{abstract}
In this work, we analyze a scalar field model which gives rise to stable bound states in field space characterized by nonzero motion that breaks the underlying time translation symmetry of its Hamiltonian, known as time crystals. We demonstrate that an ideal fluid made up of these time crystals behaves as phantom dark energy characterized by an equation of state \( w < -1 \), speed of sound squared \( c_s^2 \geq 0 \), and nonnegative energy density \( \rho \geq 0 \). 
\end{abstract}

\maketitle
\flushbottom


\section{Introduction}

Non-canonical scalar field models have been extensively studied in cosmology due to their rich dynamics and potential to explain dark energy and inflationary phenomena~\cite{ArmendarizPicon2000,Chiba2000,ArmendarizPicon2001, Melchiorri_2003}. These models exhibit interesting behaviors that besides phantom regimes in cosmology \cite{Melchiorri_2003}, lead to non-trivial stability properties in condensed matter systems ~\cite{Wilczek2012,Shapere2012}.

A representative class of models with non-canonical kinetic terms give rise to the formation of crystalline type structures which oscillate in time rather than space, known as time crystals. They arise whenever a system described by a time- independent Hamiltonian has stable bound state solutions that spontaneously break the time translation symmetry of its Hamiltonian and reduce its time translation symmetry to a discreet subgroup that is periodic in time. In this work, we investigate the time crystals model proposed in ~\cite{Shapere2012, Wilczek2012} and show that it behaves as a ideal fluid with a phantom dark energy equation of state. 

Starting our investigation with a general non-canonical Lagrangian, we first derive the general expressions for the energy density \( \rho \), pressure \( p \), equation-of-state parameter \( w \), and speed of sound squared \( c_s^2 \), and then apply these results to the time crystal model of ~\cite{Shapere2012}. 

A general non-canonical scalar field model is described by the Lagrangian density
\begin{equation}
\mathcal{L} = f(\phi) g(X) - V(\phi), \label{eq:lagrangian_general}
\end{equation}
where: \( \phi=\phi_{0}(t) + \delta\phi(t,x) \) is a scalar field with \(\phi_0\) the background field and \(\delta\phi\) perturbations about it. As is typical of inflationary models, the background field \(\phi_0 (t)\) is homogeneous while perturbations are functions of space and time (\(x,t\)). The kinetic term defined as \( X = \dfrac{1}{2}g_{\mu\nu} \nabla{\phi}^{\mu}\nabla^{\nu}\) reduces to \(X=\frac{1}{2}\dot{\phi}^2 >0\) for the homogeneous mode \(\phi_0\). Here, \( f(\phi) > 0 \) is a positive semi-definite function of \( \phi \); \( g(X) \) is a non trivial function of \( X \), and \( V(\phi) \) is the potential of the scalar field which we take to be a semi-definite positive function.

The conditions required for an ideal fluid with a phantom equation of state \( w < -1 \), a nonnegative energy density \( \rho \geq 0 \), and a positive speed of sound squared  \( c_s^2 \geq 0 \) are reviewed in Section II. In Section III, we apply these considerations to the specific model of time crystals of ~\cite{Shapere2012,Wilczek2012} and explore its connection to phantom dark energy. From here on, we focus on the dynamics of the homogeneous field \(\phi_0\) and drop the subscript \(_0\). The study of perturbations is reported in a companion paper ~\cite{lauraperturb}.

\section{General Formalism}

The energy-momentum tensor for a scalar field in a flat Friedmann-Robertson-Walker (FRW) universe is given by:
\begin{equation}
T_{\mu\nu} = \partial_{\mu} \phi \dfrac{\partial \mathcal{L}}{\partial (\partial^{\nu} \phi)} - g_{\mu\nu} \mathcal{L}.
\end{equation}

For a homogeneous scalar field \( \phi = \phi(t) \), the corresponding energy density \( \rho \) and pressure \( p \) of the ideal fluid are:
\begin{align}
\rho &= T^{0}_{\ 0} = \dot{\phi} \dfrac{\partial \mathcal{L}}{\partial \dot{\phi}} - \mathcal{L}, \label{eq:rho_def} \\
p &= -\dfrac{1}{3} T^{i}_{\ i} = \mathcal{L}, \label{eq:p_def}
\end{align}
where \( i \) runs over spatial indices. Hence, for an ideal fluid its pressure \(p\) and energy density \(\rho\) are identified with the Lagrangian and Hamiltonian of the scalar field respectively.

The relevant Einstein equations are the Friedmann equation and the Bianchi identity.
The Friedman equation is
\begin{equation}
(\frac{\dot{a}}{a})^{2} = \frac{8\pi G}{3} ( \rho +\rho_{m} )
\label{eq:friedmann}
\end{equation}
where \( a(t) \) is the scale factor of the FRW metric with line element \(ds^2 = -dt^2 + a(t)^2 dx_{(3)}^2\), and \(\dot{} = d/dt\). In general, we could allow for other matter and energy components \( \rho_{m} \) in the universe, not related to the non-canonical ideal fluid \(\phi \). However, the phantom dark energy component soon dominates the expansion while the matter and radiation components dilute away, rendering them insignificant. For this reason we ignore the contribution of \(\rho_m\) in the equations below.

The Bianchi identity of the fluid is
\begin{equation}
\dot{\rho} + 3 \frac{\dot{a}}{a} (\rho + p) = 0
\label{eq:bianchi}
\end{equation}

In this class of models, the non canonical conjugate momenta \(\Pi \), obtained as a derivative of \( \mathcal{L} \) with respect to \( \dot{\phi} \), is different from the kinetic momentum \( \dot{\phi} \) 
\begin{equation}
\Pi = \dfrac{\partial \mathcal{L}}{\partial \dot{\phi}} = f(\phi) g^{\prime}(X) \dfrac{\partial X}{\partial \dot{\phi}} = f(\phi) g^{\prime}(X) \dot{\phi},
\end{equation}
where \( g^{\prime}(X) = \dfrac{d g}{d X} \) and \( \dot{\phi} =  \pm \sqrt{2 X} \).

Due to the non-canonical nature of the Lagrangian, the Lagrangian and the Hamiltonian expressions below are quite different from one another. Substituting this into Eq.~\eqref{eq:rho_def} to obtain the Hamiltonian
\begin{align}
&\rho  = \mathcal{H} = \Pi \dot{\phi} - \mathcal{L} \\
&= f(\phi) \left( 2 X g^{\prime}(X) - g(X) \right) + V(\phi). \label{eq:rho_general}
\end{align}

whereas from Eq.~\eqref{eq:p_def}, we have:
\begin{equation}
p = f(\phi) g(X) - V(\phi). \label{eq:p_general}
\end{equation}

The equation-of-state parameter \( w \) of the cosmic fluid \( \phi \) is defined by:
\begin{equation}
w_{\phi} = \dfrac{p}{\rho} = \dfrac{f(\phi) g(X) - V(\phi)}{f(\phi) \left( 2 X g^{\prime}(X) - g(X) \right) + V(\phi)}. \label{eq:w_general}
\end{equation}

The speed of sound squared is given by~\cite{Garriga1999}:
\begin{equation}
c_s^2 = \dfrac{p_{,X}}{\rho_{,X}}, \label{eq:cs2_def}
\end{equation}
where \( p_{,X} = \dfrac{\partial p}{\partial X} \) and \( \rho_{,X} = \dfrac{\partial \rho}{\partial X} \), and,
\begin{align}
p_{,X} &= f(\phi) g^{\prime}(X), \label{eq:pX_general} \\
\rho_{,X} &= f(\phi) \left( g^{\prime}(X) + 2 X g^{\prime\prime}(X) \right), \label{eq:rhoX_general}
\end{align}
with \( g^{\prime\prime}(X) = \dfrac{d^2 g}{d X^2} \).

Therefore, the expression for the speed of sound squared becomes
\begin{equation}
c_s^2 = \dfrac{g^{\prime}(X)}{g^{\prime}(X) + 2 X g^{\prime\prime}(X)}. \label{eq:cs2_general}
\end{equation}

\subsection{Conditions for Stable Phantom Behavior with Positive Energy density}

A phantom dark energy fluid has an equation-of-state parameter \( w_{\phi} < -1 \). We demand that the energy density of the phantom fluid given by Eqn.~\eqref{eq:rho_general} is positive, and explore the field range under which the three quantities of interest simultaneously satisfy their constraints (\(\rho\ge 0, c_{s}^{2} \ge 0, w_{\phi} \le -1\)) .

Given the requirement that \(\rho\ge 0\), from Eqn.~\eqref{eq:w_general} the condition \( w_{\phi} < -1 \) is equivalent to
\begin{equation}
\rho + p < 0. \label{eq:w_lt_minus1_condition}
\end{equation}

Using Eqns.~\eqref{eq:rho_general} and \eqref{eq:p_general},
the condition \( w_{\phi} < -1 \) then implies
\begin{equation}
f(\phi) X g^{\prime}(X) < 0. \label{eq:phantom_condition}
\end{equation}

Since \( f(\phi) > 0 \) and \( X > 0 \), the condition for a phantom dark energy equation of state further simplifies to
\begin{equation}
g^{\prime}(X) < 0. \label{eq:phantom_condition_simplified}
\end{equation}

Note that the conditions for phantom behavior depends solely on the expression of the kinetic term \(g(X)\) and are completely independent of the type of potential \( V(\phi) \) chosen. 

The requirement \( \rho \geq 0 \) is equivalent to
\begin{equation}
f(\phi) \left( 2 X g^{\prime}(X) - g(X) \right) + V(\phi) \geq 0. \label{eq:rho_nonnegative_condition}
\end{equation}

Since \( f(\phi) > 0 \), it can be further simplified as:
\begin{equation}
2 X g^{\prime}(X) - g(X) + \dfrac{V(\phi)}{f(\phi)} \geq 0. \label{eq:rho_nonnegative_simplified}
\end{equation}
an inequality which can be solved, given a specific expression of \(g(X)\) and \(V(\phi)\). The \(\rho\ge 0\) requirement of Eqn.~\eqref{eq:rho_nonnegative_simplified} depends on the specific forms of \( g(X) \), \( V(\phi) \), and \( f(\phi) \). 

However, given that \( g'(X) \le 0 \) for phantom dark energy ideal fluids, and further restrictions that we will discuss below in the context of time crystals, that limit the field from exploring a wide range of the field space and confine it to the crystal region, as well as the coupling of the field to gravity on an expanding FRW background, make the dependence on the potential energy term sufficiently weak to be accommodated by a large class of generic models. Potentials which satisfy the positive energy condition would only be weakly tuned.

If the speed of sound squared were negative we would be looking at a tachyonic fluid which is wildly unstable.
Therefore, requiring the speed of sound squared of the fluid to be positive or zero provides the regions of stability in the field space.

The speed of sound squared is given by Eq.~\eqref{eq:cs2_general}. Our third requirement that \( c_s^2 \geq 0 \) leads to the following constraint on derivatives of \(g(X)\)

\begin{equation}
g^{\prime}(X) \left[ g^{\prime}(X) + 2 X g^{\prime\prime}(X) \right] \geq 0. \label{eq:cs2_positive_condition}
\end{equation}

This inequality depends on the signs of \( g^{\prime}(X) \) and \( g^{\prime}(X) + 2 X g^{\prime\prime}(X) \). But the equation of state of the phantom fluid already imposed the condition that \( g'(X) \le 0 \) , which means that \(c_s^{2} \geq 0\) can only be satisfied for 
\begin{equation}
2 X g'' (X) + g'(X) \le 0 .
\label{eq:gcondition}
\end{equation}
Combining the three constraints (\( c_s^2 \geq 0 \), \( w < -1 \), \( \rho \geq 0 \)) for the phantom regime, we now apply them to the time crystal model given in \cite{Shapere2012}. We show that the time crystal model of ~\cite{Shapere2012, Wilczek2012} gives rise to phantom dark energy and the range of \( X \le X_t\) inside the crystal boundaries satisfy all three constraints. To demonstrate this, it requires solving the inequalities described in this section simultaneously. Notice that, in contrast to k-essence models where the field can explore a large range in field space, the case of time crystals the field becomes more intriguing since the field is confined to a small range within the boundaries of the crystal, and forbidden to climb out of the crystal range.

\section{Application of Phantom Dark Energy Conditions to the Time Crystal Model}

Time crystals were first discovered in \cite{Wilczek2012, Shapere2012}. Although their discovery is relatively recent, there already exists a vast literature on the topic of time crystals and experiments set up for observing them in a lab, reviewed in \cite{timecrystalreview}. 

In this work, we take the kinetic term \(g(X)\) to be that of the time crystal model of \cite{Shapere2012} and show that it produces phantom dark energy. However, the formalism presented here can be extended to more complex time crystal models. 

\subsection{A Time Crystals Lagrangian}

Consider the time crystal Lagrangian density of ~\cite{Shapere2012},
\begin{equation}
\mathcal{L} = -\kappa_2 X + \lambda_2 X^2 - V(\phi), \label{eq:tc_lagrangian}
\end{equation}
where \( \kappa_2 > 0 \) and \( \lambda_2 > 0 \) are constants, \(f(\phi) =1\), and \( V(\phi) \) is a positive semi definite potential. The non canonical kinetic term \( g(X) \) of the previous section is 
\begin{equation}
g(X) = -\kappa_{2} X + \lambda_{2} X^2
\end{equation}

Applying the results of the previous section, the Hamiltonian obtained from this Lagrangian, is
\begin{equation}
\mathcal{H} = -\kappa_2 X + 3 \lambda_2  X^2 + V(\phi)
\label{eq:tc_hamiltonian}
\end{equation}
which can be rewritten as 
\begin{equation} 
\mathcal{H} = 3 \lambda_2 (X - X_t)^2 + (V(\phi) - \frac{\kappa_{2}^2}{12\lambda_{2}})
\label{eq:newhamiltonian}
\end{equation}
where
\begin{equation}
\mathcal{H} + \mathcal{L} = \Pi \dot{\phi}    
\label{eq:lh}
\end{equation}
and conjugate momentum,
\begin{align}
&\Pi = \frac{\partial \mathcal{L}}{\partial \dot{\phi}} = \dot{\phi} (-\kappa_{2} +2\lambda_{2} X)\\
&= \dot{\phi}(- \kappa_{2} + \lambda_{2} \dot{\phi}^2)
\label{eq:momenta}
\end{align}

The boundary \(X_t =\kappa_2 /6\lambda_2 \) defines the turning points of the bound orbit of the time crystal that minimize kinetic energy \(\frac{\partial\mathcal{H}}{\partial X}|_{X=X_t}=0\), corresponding to \( \dot\phi_{t} = \pm (\kappa_2 / 3\lambda_2)^{1/2} \ne 0\). Note that there is motion \( X\ne 0 \) along the orbit \(X_t\) which gives rise to the spontaneous breaking of the time translation symmetry of the Hamiltonian as is apparent from Eqn.~\eqref{eq:newhamiltonian}. The effective potential \(W(\phi)\) in the expression for the Hamiltonian, Eqn.~\eqref{eq:newhamiltonian}, is given by
\begin{equation}
W(\phi) = V(\phi) - \kappa_{2}^2 /12\lambda_2 = V(\phi) - \Lambda_T
\label{eq:lambda}
\end{equation}

 Rewriting the energy density in this manner, illustrates the fact that the crystal contributes an overall constant negative energy term \( \Lambda_T \) in Eqn.~\eqref{eq:lambda} that contributes to the effective potential of the energy expression of Eqn.~\eqref{eq:newhamiltonian}. The term \(\Lambda_T\) which appears naturally from time crystals, can offset a large positive vacuum energy and alleviate the notorious fine tuning of problem of vacuum energy (by \(123-\) orders of magnitude), through a far less tuned choice of its parameters \(\lambda_{2}, \kappa_{2}\). The parameters \(\lambda_{2}, \kappa_{2}\) which enter in a ratio, can take reasonable values and still be such that their combination in  the time crystal contribution to vacuum energy \(\Lambda_T\) is large or small enough to absorb the background's positive vacuum energy. 

Furthermore, we can also rewrite the Lagrangian for this model as
\begin{equation}
\mathcal{L} = \lambda_2 (X - X_d)^2 - ( V(\phi) +\frac{\kappa_{2}^2}{4\lambda_{2}} )
\label{eq:newlagrangian}
\end{equation}

where \(X_d\) is defined as the locus of the Lagrangian \(\frac{\partial\mathcal{L}}{\partial X} |_{X=X_d} =0\).
As we will see below, by writing the Hamiltonian and Lagrangian in this manner, both points \(X_{t}, X_{d}\) will play an important role in the dynamics of the fluid.

For the ideal fluid with a stress energy tensor,
\begin{equation}
    T_{\mu\nu} = (\rho + p) u_{\mu}u_{\nu} - p g_{\mu\nu}
\end{equation}
with \( u_{\mu}\) the unit 4-vector normalized to unity, whereas the pressure density of the fluid is given by the Lagrangian Eq.~\eqref{eq:newlagrangian} of the scalar field, \(p = \mathcal{L}\), and the energy density of the fluid is given by the Hamiltonian expression \(\rho = \mathcal{H}\) of Eq. \eqref{eq:newhamiltonian}. 
With this notation, we can rewrite the energy density of the fluid:
\begin{equation}
\rho = 3\lambda_{2} (X- X_{t})^2 + W(\phi)
\label{eq:tc_simplifiedenergy}
\end{equation}

Note that while the non-canonical form of the kinetic term of the time crystal contributes a term \(\Lambda_t\) which reduces the total energy of the ideal fluid by a finite negative constant energy \(- \Lambda_T = - \frac{\kappa_2^{2}}{12\lambda_2}\), meanwhile the correction to the pressure of the ideal fluid increases the pressure by a positive constant amount \(p_t = 3\Lambda_T = + \kappa_{2}^{2} /4\lambda_{2} \). The pressure correction term for the effective pressure is positive and three times larger than the energy correction in the effective potential, typical of a solids equation of state.

From Lagrangian equations of motion we have
\begin{equation}
\frac{1}{a(t)^3}\frac{d a(t)^{3} \Pi(X)}{dt}= \frac{\partial\mathcal{L}}{\partial\phi}
\label{eq:eom1}
\end{equation}
which, when expanded reads \(\frac{\partial\mathcal{L}}{\partial\phi} =3 H \Pi + \dot\Pi\), with \(H =  \frac{\dot{a}}{a}\) is the Hubble parameter. The equation of motion can be written specifically in terms of the field, by factoring out \(\rho_X\) and using the definition of the speed of sound, as
\begin{equation}
    \ddot{\phi} + 3 H c_{s}^{2} \dot{\phi} + \frac{1}{\rho_X}\frac{\partial V(\phi)}{\partial \phi}=0
    \label{eq:fieldeom}
\end{equation}

The equations of motion are directly related to three points and regions of interest in the field space of the time crystals ideal fluid, that show up as \(X_t\) and \( X_d \) in Eqns. ~\eqref{eq:newhamiltonian} and ~\eqref{eq:newlagrangian}.

The  set of Einstein's equations, e.g. \(H\simeq \sqrt{\rho}\), coupled to the field equation of motion Eqn.~\eqref{eq:fieldeom}, and the perturbations equation ~\cite{lauraperturb} can be solved numerically and we will report these results in an upcoming paper. However, since the Hubble parameter grows and the scale factor diverges within a finite time interval, according to Eqn.~\eqref{eq:eom1} the conjugate momentum is driven to zero as \(\Pi \simeq \frac{1}{a^3}\) during the same time interval \(t\simeq t_{0} w/(1+w)\) with \(t_0\) the present time ~\cite{caldwell},  as \(a(t)\rightarrow \infty\).

The turning points can be obtained by minimizing kinetic energy or equivalently from the cusps of the conjugate momentum. The cusps of the conjugate momentum \(\Pi(X)\) indicate critical points in the fields space because, as can be seen by a direct inspection, the speed of sound diverges there \(c_{s}^{2} ->\infty\). These cusps are found by
\begin{equation}
    \frac{\partial \Pi}{\partial \dot{\phi}} =0
\end{equation}
solutions of which occur exactly at the turning points of the time crystal \(\dot{\phi}_t = \pm \sqrt{\kappa_{2}/3\lambda{2}}\) or \(X=X_t\). The latter, in combination with the fact that the speed of sound squared diverges at the boundary, indicate that these bound structures, the time crystals, are stable against perturbations ~\cite{lauraperturb} and the field is confined to the time crystal region \(X\le X_t\) where the boundary of the crystal behaves as a hard wall. The latter should be contrasted to k-essence models where the field explores a large range in field space.
Hence, the time crystal bound state is a state of minimum kinetic energy, despite that the underlying time translation symmetry is broken and there is motion \( X \ne 0\) along the orbit.
The meaning of (\(X_t, \pm\phi_t\)) then becomes clear as the turning points that define the orbit of the bound state, but which, due to motion break time translation symmetry of the Hamiltonian and become time crystals.  

Another point of interest is the one where the group velocity of the field vanishes and the kinetic energy is once again minimized, the locus of the Lagrangian \(\Pi = \frac{\mathcal{L}}{\partial\dot{\phi}} = 0\). As can be seen from Eq.~\eqref{eq:momenta}, this occurs at \(X= X_d = \frac{\kappa_2}{2\lambda_2}\) or equivalently along the orbit defined by the special points \(\phi_d = \pm \sqrt{\kappa_{2}/\lambda_2}\) that enter the Lagrangian expression, Eqn.~\eqref{eq:newlagrangian}. However these points lie outside the crystal's allowed field range, \(X_d > X_t\). Due to the fact that the boundary of the crystal \(X=X_t\) is a hard wall type with a diverging speed of sound squared, the field cannot access regions \(X>X_t\) beyond the crystal boundary, therefore it can not reach \(X_d\).

\subsection{Phantom Dark Energy from the Time Crystal model}

Let us now investigate the relation between the time crystal model described in the previous section to phantom dark energy fluids with (\(w < -1\), \(c_{s}^{2} >0\), \(\rho >0 )\).  
The first and second derivatives of \( g(X) \) are:
\begin{align}
g^{\prime}(X) &= -\kappa_2 + 2 \lambda_2 X, \label{eq:g_prime_tc} \\
g^{\prime\prime}(X) &= 2 \lambda_2. \label{eq:g_double_prime_tc}
\end{align}

The condition for Phantom dark energy behavior, (\( w < -1 \)) provided that \(\rho\ge 0\),  Eq.~\eqref{eq:phantom_condition_simplified}, is:
\(g^{\prime}(X) < 0 \).
Substituting \( g^{\prime}(X) \) from Eq.~\eqref{eq:g_prime_tc} gives 
\begin{equation}
X < \dfrac{\kappa_2}{2 \lambda_2}. \label{eq:X_phantom_tc}
\end{equation}

Hence, phantom behavior occurs anywhere in the region below the point \(X \le X_d \) where group velocity of the field and conjugate momentum \(\Pi\) vanish:
\begin{equation}
0 < X < \dfrac{\kappa_2}{2 \lambda_2}.
\end{equation}

The requirement for a nonnegative energy density (\( \rho \geq 0 \)), with \( f(\phi) = 1 \) is satisfied by:
\begin{equation}
-\kappa_2 X + 3 \lambda_2 X^2 + V(\phi) \geq 0. \label{eq:rho_positive_condition_tc}
\end{equation}
which, for any specified potential, is straightforward to solve as a quadratic inequality.
For potentials that satisfy \(W(\phi) \geq 0 \), equivalently \( V(\phi)\geq \frac{\kappa_2^{2}}{12 \lambda_2}\), the energy density is positive everywhere, as it's clear from Eqn. ~\eqref{eq:newhamiltonian}. 

Given the above bounds on \(g(X), W(\phi)\) it is easy to see that the energy density is always positive in the region of interest where the ideal fluid as phantom behavior, which included the time crystal \(X\le X_t\).

The equation of state parameter \(w_\phi\) for the ideal fluid described by the time crystals model is of a phantom type,  
\begin{equation}
w_{\phi} = \frac{p}{\rho} = \frac{-\kappa_{2} X + \lambda_{2} X^{2} - V(\phi)}{-\kappa_{2} X + 3\lambda_{2} X^{2} + V(\phi)} \le -1
\end{equation}
 for all \(X\le X_d\).
 
Finally let's check the condition for stability (\( c_s^2 \geq 0 \)) of the phantom fluid that enters the field equation, Eqn.~\eqref{eq:fieldeom}. At this point, the behavior of the model becomes interesting.

Using Eq.~\eqref{eq:cs2_general}, we have:

\begin{equation}
c_s^2 = \dfrac{ -\kappa_2 + 2 \lambda_2 X }{ -\kappa_2 + 6 \lambda_2 X }. \label{eq:cs2_tc}
\end{equation}

The requirement \( c_s^2 \geq 0 \), implies:
\begin{equation}
\left( -\kappa_2 + 2 \lambda_2 X \right) \left( -\kappa_2 + 6 \lambda_2 X \right) \geq 0. \label{eq:cs2_positive_tc}
\end{equation}
which varies across the \(X-\)space.

The variation of the behavior of the field and of the speed of sound squared across the whole range of (\(\phi, X\)) divides this range into three regions, as summarized below.
Given:
\begin{align}
p_X &= -\kappa_2 + 2 \lambda_2 X, \\
\rho_X &= -\kappa_2 + 6 \lambda_2 X.
\end{align}

The three regions are:

1. The time crystal region: \( X \leq \dfrac{\kappa_2}{6 \lambda_2} \)
   \begin{itemize}
   \item \( p_X < 0 \),
   \item \( \rho_X < 0 \),
   \item \( c_s^2 > 0, w_{\phi}<-1, \rho>0 \).
   \end{itemize}

2. The forbidden region bounded below by the orbit of the time crystal \(X_t\) and above by the locus of the Lagrangian \(X_d\) respectively: \( \dfrac{\kappa_2}{6 \lambda_2} < X < \dfrac{\kappa_2}{2 \lambda_2} \)
   \begin{itemize}
   \item \( p_X < 0 \),
   \item \( \rho_X > 0 \),
   \item \( c_s^2 < 0,\) (unstable) goes through zero at \(X=X_0\), and \(\rho>0, w_{\phi}<-1\).
   \end{itemize}

3. The quintessence region: \(X \geq X_{d} = \dfrac{\kappa_2}{2 \lambda_2} \)
   \begin{itemize}
   \item \( p_X \geq 0 \),
   \item \( \rho_X \geq 0 \),
   \item \( c_s^2 \geq 0, w_{\phi} >-1, \rho > 0 \).
   \end{itemize}

There is an additional point of interest in the forbidden region \(\it{2}\), where the speed of sound goes through zero again and halfway in this region it switches from negative to positive indicating stability, at \(X= X_0 = \dfrac{\kappa_2}{4\lambda_2} = X_{d}/2\) found by the condition of Section \(\it{1}\): \(2 X 
 g"(X) \leq - g'(X)\). In the forbidden region  \(X_{t} \le X \le X_0\) the momentum \(\Pi\) obtained by the quadratic equation above \(\rho \geq 0\) is imaginary and the field solution is a decaying mode. 
Therefore, stability (\( c_s^2 \geq 0 \)) occurs only when the field is confined in the time crystal region or when if it were to start at high conjugate moment for \(X\geq X_d\):
\begin{equation}
X \leq \dfrac{\kappa_2}{6 \lambda_2} \quad \text{or} \quad X \geq \dfrac{\kappa_2}{2 \lambda_2}. \label{eq:stability_condition_tc}
\end{equation}

However, the unstable region above \(X_t \) is inaccessible to the field since time crystals are in stable bound states and the speed of sound squared diverges at the boundary \(X=X_t\), as given by cusps in the momentum space. These imply that the field is confined in the crystal region and it would require infinite work done to it to jump over the crystal boundary and to break the crystal structure. Even if the field were to tunnel through \(X_t\) boundary with a small probability it would die out to zero quickly since the momentum becomes imaginary and the field solution is a decaying mode in the forbidden region. In other words, these crystals are very stable structures, and the field is confined within the boundaries of the crystal.

The only region accessible to the field is the time crystal interval that simultaneously satisfies all the conditions for the phantom dark energy which is the time crystal region: \( 0 < X \leq \dfrac{\kappa_2}{6 \lambda_2} \), provided \( W(\phi) \geq 0 \).
In this region, we find the following.

 \( w_{\phi} < -1 \) (phantom behavior),
 \( c_s^2 \geq 0 \) (stability),
\( \rho \geq 0 \) if the potential \( V(\phi) \) is sufficiently large to ensure \( \rho \geq 0 \).The latter is automatically satisfied throughout the field space if we require \(W(\phi) \geq 0\), which can be easily achieved by multiplying the potential \(V(\phi)\) by an overall constant factor without changing or tuning its functional form. Despite that \(p_{X}\) and \(\rho_X\) are negative within the crystal boundaries, the important factor for the stability of the field is their ratio which gives the speed of sound squared be positive, \(c_{s}^{2}= \frac{p_X}{\rho_X} >0 \). As can be seen from the field equation in an expanding FRW universe, Eqn.~\eqref{eq:fieldeom}, the Hubble drag term multiplied by the speed of sound overwhelmingly dominates over the potential term. The only effect of \(\rho_X < 0\) is to invert the sign of the potential, a term which is insignificant relative to the growing Hubble drag term, ( since \(H \simeq a(t)^{\frac{-3}{2(1+w)}}, c_{s}^{2}\ge 1\). (This stability remains true for the field perturbations also, as studied in ~\cite{lauraperturb}.)

For \( X \geq \dfrac{\kappa_2}{2 \lambda_2} \), stability is maintained and \( \rho \geq 0 \), but \( w_{\phi} > -1 \), so the phantom behavior is lost and replace by a quintessence type dynamics. However, if the field is bound below \(X_d\) in the time crystal region it would not be possible for the field to break away from its bound state and suddenly jump over to \(X=X_d\) which is the boundary of the quintessence region, with the momentum \(\Pi_d =0 \) and the group velocity are zero at that initial boundary. The field could not access the region \(X > X_t \).

The table below \ref{tab:critical_points} classify the three regions of interest in field space, and summarizes these results, detailing the range of \( X \), energy density, equation-of-state parameter, and sound speed squared \( c_s^2 \) for all the field space regions, as discussed above.

\begin{table*}[t]
    \centering
    \caption{Classification of Key Points and Regions in the Time Crystal Scalar Field Model}
    \label{tab:critical_points}
    \scalebox{0.8}{
    \begin{tabular}{|c|c|c|c|c|}
        \hline
        \textbf{Region} & \textbf{Range of \( X \)} & \boldmath$\rho$ & \boldmath$w_\phi$ & \boldmath$c_s^2$ \\
        \hline
        Trivial canonical vacuum energy & \( X = 0 \) & \( \rho = V(\phi) > 0 \) & \( w_\phi = -1 \) & \( c_s^2 = 1 \) \\
        \hline
        Time Crystal Region & \( 0 < X < X_t = \frac{\kappa_2}{6\lambda_2} \) & \( \rho > 0 \) & \( w_\phi < -1 \) & \( c_s^2 > 1 \) \\
        \hline
        Time Crystal Boundary & \( \lim_{X \rightarrow X_t^-= \frac{\kappa_2}{6\lambda_2}} \) & \( \rho = -\frac{\kappa_2^2}{12\lambda_2} + V(\phi) > 0 \) & \( w_\phi\rightarrow w_\phi^{\rm min} < -1 \) & \(c_s^2 \to \infty \) \\
        \hline
        Forbidden Region & \( X_t < X < X_d = \frac{\kappa_2}{2\lambda_2} \) & \( \rho > 0 \) & \( w_\phi < -1 \) & \( c_s^2 < 0 \),at\(X=X_d /2\) \(c_s^2=0\), then \(c_s^2>0\) \\
        \hline
        Boundary of quintessence  & \( \Pi=0, X = X_d = \frac{\kappa_2}{2\lambda_2} \) & \( \rho = \frac{\kappa_2^2}{4\lambda_2} + V(\phi) > 0 \) & \( w_\phi = -1 \) & \( c_s^2 = 0 \) \\
        \hline
        quintessence  & \( X > X_d = \frac{\kappa_2}{2\lambda_2} \) & \( \rho = \frac{\kappa_2^2}{4\lambda_2} + V(\phi) > 0 \) & \( w_\phi > -1 \) & \( 0<c_s^2<1 \) \\
        \hline
    \end{tabular}
    }
\end{table*}

The Hamiltonian is a single-valued function at the trivial point \( X = 0 \), the equation of state \(w_{\phi}=-1\) is that of a pure positive cosmological constant and the speed of sound \(c_{s}^{2} =1\) indicates a canonical ideal fluid.

In the region between zero and the turning points for the orbit \( 0 < X < X_t = \frac{\kappa_2}{6\lambda_2} \) we have the time crystal which is a stable bound state, for the reasons discussed above. The Hamiltonian is a double-valued function of \(\dot{\phi}\). 
The energy density remains positive:
\begin{equation}
    \rho_\phi = -\kappa_2 X + 3\lambda_2 X^2 + V(\phi) > 0.
\end{equation}
The equation-of-state parameter satisfies:
\begin{equation}
    w_\phi < -1,
\end{equation}
characterizing phantom-like behavior. The sound speed squared is positive:
\begin{equation}
    c_s^2 = \frac{ -\kappa_2 + 2\lambda_2 X }{ -\kappa_2 + 6\lambda_2 X } > 0.
\end{equation}
This ensures stable perturbations within this interval.
At the boundary \( X = X_t\) the energy density reaches:
\begin{equation}
    \rho = -\frac{\kappa_2^2}{12\lambda_2} + V(\phi) > 0.
\end{equation}
Here, the equation-of-state parameter \( w_\phi \) attains its minimum value within the phantom regime (\( w_\phi < -1 \)). The sound speed squared \( c_s^2 \rightarrow \infty\) at the crystal's boundary, indicating a stable time crystal structure which confines the field within. 

The field would require an infinitely large pressure gradient, or force, to break out of the crystal boundary \(X_t\). In a sense, this boundary acts as a hard wall potential for the forbidden region \(X > X_t\), with the field trapped on its other side \(X \leq X_t\).
Hence, the region \( X_t < X < X_d = \frac{\kappa_2}{2\lambda_2} \) is inaccessible to the field.

The speed of sound squared being greater than one within the crystal region, is typical of crystalline structures.  Its contribution to the metric perturbation equation is to dampens perturbations ~\cite{caldwell} that would contribute to cold dark matter, thereby offering a smoking gun for testing this model.

In the region above the time crystal orbit,  the scalar field continues to exhibit phantom-like behavior with
\begin{equation}
    w_{\phi} < -1.
\end{equation}
The energy density remains positive:
\begin{equation}
    \rho > 0.
\end{equation}
However, the sound speed squared becomes negative and the momentum imaginary
\begin{equation}
    c_s^2 = \frac{ -\kappa_2 + 2\lambda_2 X }{ -\kappa_2 + 6\lambda_2 X } < 0,
\end{equation}
The speed of sound squared goes from negative values through zero halfway in the region at \(X =X_0 = X_{d}/2\) and then it switches to positive values.

At the next boundary \( X = X_d \) where the conjugate momentum \(\Pi\) and group velocity start from zero, the energy density is:
\begin{equation}
    \rho_\phi = \frac{\kappa_2^2}{4\lambda_2} + V(\phi) > 0.
\end{equation}
The equation-of-state parameter returns to \(-1\):
\begin{equation}
    w_\phi = -1,
\end{equation}
resembling a cosmological constant. The sound speed squared \( c_s^2 \) is unity:
\begin{equation}
    c_s^2 = 1.
\end{equation}
This ensures stable perturbations at this critical point.

Above \(X_d\) in the region where \( X_d < X \), the conjugate momentum \(\Pi\) starts increasing and the scalar field transitions to quintessence-like behavior with:
\begin{equation}
    -1 < w_\phi < 0.
\end{equation}
The energy density remains positive:
\begin{equation}
    \rho_\phi > 0.
\end{equation}
The sound speed squared satisfies:
\begin{equation}
    0 < c_s^2 < 1,
\end{equation}
indicating stable perturbations within the quintessence regime.
However, for the reasons discussed above, regions above the boundary of the time crystal with \(X\ >X_t\) are not accessible due to the hard wall at the boundary \(X_t\) of the crystal where \(c_{s}^2 \rightarrow \infty\). Perturbations around the crystal solution, (\(\ddot{\delta\phi} +\frac{c_{s}^{2} k^2}{a^2} + (3H + \frac{6\lambda_{2}\ddot{\phi}\dot{\phi}}{\rho_X})\dot{\delta\phi} + \frac{1}{\rho_X}\frac{\partial^2 V}{\partial\phi^2}=0 \)),
where studied in \cite{lauraperturb} and shown to be stable.  Besides, the time crystal region \(X\le X_t\) is favored as the state that minimize energy.

\section{Conclusions}
Until recently, phantom dark energy models have been studied in the context of k-essence fluids with non-canonical kinetic energies.
In this work, we investigated the time crystal model proposed in ~\cite{Shapere2012, Wilczek2012} and showed that in the context of cosmology, it gives rise to phantom dark energy, with an equation of state \(w_{\phi} \le -1\), a positive speed of sound squared \(c_{s}^{2} \ge 0\), and positive energy density \(\rho \ge 0\), which drives the expansion of the universe towards the Big Rip. While both, k-essence models and the time crystal model discussed here,  have in common non-canonical kinetic terms, in the latter case we are dealing with a collection of crystalline structures that break time symmetry and behave as a phantom fluid instead of matter. Furthermore, in contrast to k-essence models where the field explores a large volume in field space, in time crystals the field is confined within the boundary of the crystal whereby the rest of the field space is inaccessible by a hard wall. 

The possibility of ghost or gradient instabilities are often a cause of concern in k-essence models. These instabilities on the field and its perturbations do not arise in the present model. The time crystals are on an expanding FRW background where the ever increasing Hubble drag term, multiplied by a crystal's positive speed of sound squared, dominates over the potential term.  

Furthermore, the speed of sound in the crystal region \(X\le X_t\) is such that \( c_{s}^{2}\ge 1\) typical of crystalline structures. Observationally this sound speed would modify the Newtonian potential in a unique way, which could provide a way of testing this model. The couple equations of metric perturbations and field equations can only be solved numerically and we will report this results in an upcoming paper.

The speed of sound squared diverges at the boundary of the time crystal in field space \(X_t\) which implies that these, and their perturbations,  are very stable structures bound by a hard wall in field space.
Yet, not only is the field confined to remain in the crystal region \(X\le X_t\) but dynamics favors it to naturally be found there since the crystal state minimizes energy.

Finally, the crystalline structure is time, which behaves as a phantom dark energy ideal fluid, naturally contribute an overall constant negative vacuum energy, \(\Lambda_T\). Without the need of fine-tuning that is typical of the cosmological constant problem, this contribution to vacuum energy can offset a positive cosmological constant, with a reasonable, less-tuned, choice for the values of the parameters \( (\lambda_{2}, \kappa_{2} ) \) whose ratio determines the magnitude of  \(\Lambda_T \).

\noindent{ \textit{Acknowledgment}: L. Mersini-Houghton is grateful to Y. J. Ng, O. Akarsu, E. Di Valentino for useful discussions, and to the Klingsberg foundation and the Bahnson fund for their support.}

\end{document}